\documentclass[12pt]{article}
\usepackage{amssymb,amsfonts}
\usepackage{latexsym}
\usepackage{amsmath}
\usepackage{slashed}
\usepackage{graphicx}
\usepackage{hyperref}
\usepackage{array}   % for \newcolumntype macro
\usepackage{ulem}
\usepackage{epsfig}
\usepackage{psfrag}
\usepackage[margin=1cm]{caption}

\usepackage{color}
\usepackage{cite}
\usepackage[lmargin=2cm,rmargin=2cm,bmargin=3cm,tmargin=2cm]{geometry}

\makeatletter\@addtoreset{equation}{section}\makeatother

\DeclareMathOperator{\Li}{Li}

\DeclareMathOperator{\sign}{sign}
\DeclareMathOperator{\ch}{ch}
\DeclareMathOperator{\sh}{sh}

\def\bS {\mathbb{S}}

\newcommand{\beq}{\begin{equation}}
\newcommand{\eeq}{\end{equation}}
\newcommand{\bal}{\begin{equation}\begin{aligned}}
\newcommand{\eal}{\end{aligned}\end{equation}}

\newcommand{\eqn}[1]{(\ref{#1})}

\newcommand{\cC}{{\mathcal C}}

\newcommand{\cN}{{\mathcal N}}

\newcommand{\cO}{{\mathcal O}}

\newcommand{\address}[1]{\vbox{\center\em#1}}
\renewcommand{\title}[1]{\vbox{\center\huge{#1}}\vspace{5mm}}

\graphicspath{{figures/}}

\begin{document}
\begin{titlepage}
\begin{center}
\phantom{xx}

\vspace{20mm}

\title{More Large $N$ limits of 3d gauge theories}

\vspace{10mm}

\renewcommand{\thefootnote}{$\alph{footnote}$}

Louise Anderson\footnote{\href{mailto:louise.m.a.anderson@imperial.ac.uk}
{\tt louise.m.a.anderson@imperial.ac.uk}},
\address{The Blackett Laboratory, Imperial College London,\\
Prince Consort Road, London SW7 2AZ, United Kingdom}
\vskip6mm
and
Nadav Drukker\footnote{\href{mailto:nadav.drukker@gmail.com}
{\tt nadav.drukker@gmail.com}}
\address{Department of Mathematics, King's College London,
\\
The Strand, London WC2R 2LS, United Kingdom
}

\renewcommand{\thefootnote}{\arabic{footnote}}
\setcounter{footnote}{0}

\end{center}

\vspace{8mm}
\abstract{
\normalsize{
\noindent
In this paper we study the large $N$ solution to matrix models describing 
the partition functions of 3d supersymmetric gauge theories on $\bS^3$. The model 
we focus on has a single $U(N)$ gauge group and fundamental fields, whose number 
scales with $N$, also known as the Veneziano limit. The novel point in our model is 
our choice of masses for the fundamental fields. Instead of vanishing or fixed masses, 
we consider a linear distribution. We show that the model can still be solved by 
standard large $N$ techniques and explore the different phases of the model. 
We also comment about other natural mass distributions.
}}
\vfill

\end{titlepage}

%\tableofcontents

\section{Introduction}

The tools of supersymmetric localization \cite{Pestun2012} 
have revolutionized the study of supersymmetric 
field theories. They allow one, under very restrictive assumptions, to reduce the infinite 
dimensional path integral of a field theory on certain manifolds to a finite dimensional 
integral. One of the most important applications has been to test dualities, as the partition 
functions of two dual models should be the same. Most often the resulting integral expressions 
are not identical, but are equal after using some integral identities.

This is a great achievement, but it is not the same as actually calculating the partition function. 
In the most celebrated case, that of 4d $\cN=2$ theories on $\bS^4$, the resulting integral is 
extremely complicated - it is equivalent to a calculation in a 2d CFT, via the AGT 
correspondence \cite{Alday:2009aq,Wyllard:2009hg} 
- but explicit evaluation of the integrals is not available even for $SU(2)$ groups 
(Liouville theory). 

In this paper we study the matrix model for 3d theories on $\bS^3$ \cite{Kapustin2010}. The 
case of the
ABJ(M) model and other circular quivers has received a lot of attention and the matrix model 
has been solved for many such cases either in the large $N$ limit, sometimes 
to all orders in $1/N$ and in a select few cases exactly 
\cite{Marino:2009jd, Drukker:2010nc, Marino2012, Codesido:2014oua}. For linear quivers 
the case of small $N$ (with no Chern-Simons term) can be solved completely 
explicitly by rather elementary integration \cite{Benvenuti:2011ga}, 
but the large $N$ limit is more complicated.

We study a model with $\cN=3$ SUSY comprising of a single $U(N)$ node with $K$ 
flavours. The partition function of the theory on $\bS^3$ is given by the matrix model 
\cite{Kapustin:2010xq}
\beq
\label{original-MM}
Z=\frac{1}{N!} \int d^N z
\frac{\prod_{i <j} \sh^2 (z_i-z_j)}
{\prod_{i =1}^N \prod_{k=1}^{K} \ch (z_i - m_k)}
e^{2 \pi i \zeta \sum_i z_i +\pi i\kappa\sum_i z_i^2}\,.
\eeq
where $m_i$ are $K$ arbitrary masses, $\zeta$ is the Fayet-Iliopoulos (FI) parameter 
and $\kappa$ the Chern-Simons (CS) level. The FI term can be eliminated by a 
shift of the integration variables and the masses $m_k$, 
and therefore will henceforth be ignored.

We consider this model in the large $N$ and large $K$ limit, also known as the 
Veneziano limit, which raises the question of how to choose the $K$ mass parameters. 
In \cite{Barranco:2014tla,Russo:2014bda} 
this model was studied with all vanishing masses or taking 
two values 
$\pm M$. The purpose of this note is to show that this model has a nice large $N$ 
solution also when the masses are distributed along an interval. We choose the 
linear distribution
\beq
\label{masses}
m_k=m_1+\frac{k-1}{K-1}\,\mu\,,
\eeq
and solve the model in the large $N$ limit.

The planar solution of the matrix model has the eigenvalues $z_i$ distributed along a cut 
(we only consider single cut solutions, which is consistent with numerical checks), and we 
find an interesting interplay between the distribution of eigenvalues and the mass 
distribution. In the absence of a CS term, the eigenvalues  are centered around 
the masses with two possibilities: The width of the eigenvalue distribution may 
be larger or smaller than that of the masses. The CS term provides an extra force 
attracting the eigenvalues to the origin and if the masses are not distributed 
symmetrically, this leads to further configurations with partial overlap or no overlap of the 
eigenvalues and the mass distribution.

The rest of the note is organised as follows: 
In the next section we solve the model. In Section~\ref{sec:graphs} we plot some graphs 
of eigenvalue distributions and investigate their possible forms. 
In Section~\ref{sec:decomp} we study the model in the limit of large $\bS^3$ 
radius, where it simplifies dramatically and 
where the different possible overlaps of the masses and eigenvalue density outlined above 
correspond to different phases with a rich structure of 
third order phase transitions. We conclude with a discussion.

\section{Solving the matrix model}
\label{sec:solve}

To solve the matrix model \eqn{original-MM}, we first change variables
\beq
\label{eq:exponentiated_variables}
Z_i=C\,e^{2\pi z_i}\,,
\qquad
M_k=C\,e^{2\pi m_k}\,,
\eeq
where $C$ is an arbitrary constant. 
Similarly, if the eigenvalues $z_i$ are supported on the interval $[a,b]$, 
then  the exponentiated  eigenvalues $Z_i=Ce^{2\pi z_i}$  are supported on $[A,B]$.
The partition function is now
\bal
Z&=\frac{\prod_{k=1}^KM_k^{N/2}}{(2\pi)^NN!}
\int d^N Z\,\prod_{i<j}(Z_i-Z_j)^2
\frac{\prod_{i=1}^N Z_i^{K/2-N}e^{i\frac{\kappa}{4\pi}\log^2(Z_i/C)}}
{\prod_{i=1}^N\prod_{k=1}^K(Z_i+M_k)}
\\&
=\frac{1}{N!}
\int d^N Z\,\prod_{i<j}(Z_i-Z_j)^2\,e^{-N\sum_{i=1}^N V(Z_i)}\,,
\eal
with the potential
\bal
V(Z)&=C_0
+\frac{1}{N}\sum_{k=1}^K\log(Z+M_k)
-\left(\frac{K}{2N}-1\right)
\log\frac{Z}{C}-i\frac{\kappa}{4\pi N}\log^2\frac{Z}{C}\,,
\\&\ \quad 
C_0=\frac{1}{N}\log(2\pi)-\left(\frac{K}{2N}-1\right)\log C-\frac{1}{2N}\sum_{k=1}^K\log M_k\,.
\eal

The saddle point equation for an eigenvalue $Z_i$ is
\beq
2\sum_{j\neq i}\frac{1}{Z_i-Z_j}
=NV'(Z_i)
=\sum_{k=1}^K\frac{1}{Z_i+M_k}
-\left(\frac{K}{2}-N\right)\frac{1}{Z_i}-i\frac{\kappa}{2\pi}\frac{\log(Z_i/C)}{Z_i}\,,
\eeq

We want to study this model in the large $N$ limit, where the saddle point 
approximation becomes exact, and 
we are then faced with choosing a distribution for the masses. 
As mentioned in the introduction, we focus on the simple choice of a 
constant distribution of masses between $m_1$ and $m_K$ according to \eqn{masses}.

As we take the large $N$ limit, it  is useful to introduce the Veneziano parameter 
\beq
\chi =\frac{K}{2N}\,,
\eeq
as well as the 't~Hooft coupling
\beq
\lambda=\frac{N}{\kappa}\,,
\eeq
and then take the large $N$ limit in such a way that both these parameters are kept fixed. 
Therefore, in this limit we have
\bal
\frac{1}{N}\sum_{k=1}^K\log(Z+M_k)
\quad\to\quad
&\frac{\chi}{\pi(m_K-m_1)}\int_{M_1}^{M_K}\frac{dM}{M}\log(Z+M)
\\&
=2 \chi \log Z-\frac{2 \chi}{2\pi\mu}
\left(\Li_2\left(-\frac{M_K}{Z}\right)-\Li_2\left(-\frac{M_1}{Z}\right)\right).
\eal
The force resulting from the full potential is
\beq
\label{cont-force}
-V'(Z)=\frac{1}{Z}\left(\frac{2 \chi}{2\pi\mu}\log\frac{Z+M_K}{Z+M_1}
+\frac{i}{2\pi \lambda}\log Z
-\left(\chi+1+\frac{i}{2\pi \lambda}\log C\right)\right).
\eeq

In order to find the eigenvalue distribution, we use the standard technique and introduce 
the resolvent $\omega(Z)$ defined by
\beq
\omega(Z)=\oint dZ'\frac{\rho(Z')}{Z-Z'}
\eeq
Assuming that the eigenvalues are distributed along a single cut, then for a generic 
function $V'(Z)$, the resolvent is given by
\beq
\label{eq:resolvent_general}
\omega(Z)=\frac{1}{2}\oint_{\cC}\frac{dZ'}{2\pi i}\frac{V'(Z')}{Z-Z'}
\sqrt{\frac{(Z-A)(Z-B)}{(Z'-A)(Z'-B)}}
\eeq
where $\cC$ is the path around the branch cut between $A$ and $B$. 
From this, one can then obtain the density by studying the discontinuity across the branch cut
\beq
\label{eq:density_disc}
\rho(Z)=-\frac{1}{2\pi i}\big(\omega(Z+i\epsilon)-\omega(Z-i\epsilon)\big).
\eeq

We rewrite equation \eqn{eq:resolvent_general}) as
\beq
\omega(Z)=\frac{1}{2}V'(Z)-\frac{1}{2}M(Z)\sqrt{(Z-A)(Z-B)}\,,
\eeq
where $M$ is defined by the integral over a deformation of the contour $\mathcal{C}$ 
to one that encircles $\infty$
\beq
\label{eq:M}
M(Z)=\oint_{\infty}\frac{dZ'}{2\pi i}\frac{V'(Z')}{Z'-Z}\frac{1}{\sqrt{(Z'-A)(Z'-B)}}\,.
\eeq
Our expression for the force includes poles and terms of the form $\frac{\log(Z+M)}{Z}$. 
It is convenient to consider their contribution to the resolvent independently.

Adapting the solution of the Chern-Simons matrix model 
\cite{Marino:2004eq,Marino:2011nm} to a generic logarithmic 
term
\beq
V'(Z)=\frac{\log(Z+M)}{Z}\,,
\eeq
gives
\beq
M_\text{log}(Z)
=\frac{1}{2 \pi i}\oint_{\infty}dZ'
\frac{\log \left(Z'+M\right)}{Z'(Z'-Z)\sqrt{(Z'-A)(Z'-B)}}\,.
\eeq
The contour is such that we get a contribution from the logarithmic branch cut 
and from the pole at $Z'=0$ (but not from $Z'=Z$), resulting in
\bal
\label{eq:M_log}
M_\text{log}(Z)
&=
-\int_{-\infty}^{-M} dZ'
\frac{1}{Z'(Z'-Z)\sqrt{(Z'-A) (Z'-B)}}
-\frac{\log M}{Z \sqrt{AB}}
\\ 
&=
\frac{2 }{Z \sqrt{A B}}
\log \frac{\sqrt{A}+\sqrt{B}}{\sqrt{B (A+M)}+\sqrt{A (B+M)}}
\\&\quad{}
+\frac{1}{Z \sqrt{(Z-A)(Z-B)}}
\log \frac{\left(\sqrt{(A-Z)(B+M)}-\sqrt{(B-Z)(A+M)}\right)^2}
{(M+Z) \left(\sqrt{A-Z}-\sqrt{B-Z}\right)^2}\,.
\eal

The contribution of the pole term is much simpler. For
\beq
V'(Z)=\frac{1}{Z}\,,
\eeq
we 
find
\beq
M_\text{pole}(Z)=-\frac{1}{Z \sqrt{AB}}\,.
\eeq

Combining all those and the usual Chern-Simons terms together 
gives
\begin{align}
\label{full-res}
\omega(Z)
&=\frac{1}{2Z}\left(\chi+1
+\frac{i}{2\pi \lambda}\log C\right)
\left(1+\frac{\sqrt{(A-Z) (B-Z)}}{\sqrt{AB}}\right)
\nonumber\\&\quad{}
-\frac{\sqrt{(A-Z) (B-Z)}}{Z\sqrt{AB}}
\left(\frac{\chi}{\pi\mu}
\log\frac{\sqrt{B(A+M_K)}+\sqrt{A(B+M_K)}}
{\sqrt{B(A+M_1)}+\sqrt{A(B+M_1)}}
-\frac{i}{2\pi \lambda}
\log\frac{2\sqrt{AB}}{\sqrt{A}+\sqrt{B}}
\right)
\nonumber\\
&\quad{}
+\frac{\chi}{2\pi\mu Z}
\left(
\log\frac{\left(\sqrt{(A-Z)(B+M_K)} -\sqrt{(B-Z)(A+M_K)}\right)^2}
{\left(\sqrt{(A-Z)(B+M_1)} -\sqrt{(B-Z)(A+M_1)}\right)^2}
-\log\frac{(Z+M_K)^2}{(Z+M_1)^2}
\right)
\nonumber\\&\quad{}
+\frac{i}{4\pi \lambda Z}
\log\frac{\big(\sqrt{A(Z-B)}-\sqrt{B(Z-A)}\big)^2}
{Z^2\left(\sqrt{Z-A}-\sqrt{Z-B}\right)^2}\,.
\end{align}

\subsection{Asymptotic behaviour of the resolvent}

Expanding $\omega(Z)=\omega^{(0)}+\omega^{(1)}/Z+\cO(Z^{-2})$, the requirement that
$\omega^{(0)}=0$ can be expressed as:
\beq
\label{omega0}
\chi+1
+\frac{i}{\pi \lambda }\log\frac{\sqrt{C}(\sqrt{A}+\sqrt{B})}{2\sqrt{AB}}
=
\frac{2\chi}{\pi\mu}
\log\frac{\sqrt{B(A+M_K)}+\sqrt{A(B+M_K)}}
{\sqrt{B(A+M_1})+\sqrt{A(B+M_1)}}\,.
\eeq
The condition $\omega^{(1)}=1$ can be recast as 
$\omega^{(1)}+\frac{A+B}{2} \omega^{(0)}=1$ 
which gives
\beq
\label{omega1}
-\chi+1+\frac{i}{\pi\lambda}\log\frac{\sqrt{A}+\sqrt{B}}{2\sqrt{C}}
=\frac{2\chi}{\pi\mu}\log\frac{\sqrt{A+M_K}-\sqrt{B+M_K}}{\sqrt{A+M_1}-\sqrt{B+M_1}}\,.
\eeq
For $\kappa\neq0$, \textit{i.e.} finite $\lambda$, we can choose $C$ to simplify these equations. 
For $\lambda\rightarrow \infty$, the $C$-dependence drops out.

We can use \eqn{omega0} to simplify the resolvent \eqn{full-res}
\bal
\label{eq:resolvent-simplified}
\omega(Z)
&=\frac{1}{2Z}\left(\chi+1
+\frac{i}{2\pi \lambda}\log C\right)
\\&\quad{}
+\frac{\chi}{2\pi\mu Z}
\left(
\log\frac{\left(\sqrt{(A-Z)(B+M_K)} - \sqrt{(B-Z)(A+M_K)} \right)^2}
{\left(\sqrt{(A-Z)(B+M_1)} -\sqrt{(B-Z)(A+M_1)}\right)^2}
-\log\frac{(Z+M_K)^2}{(Z+M_1)^2}
\right)
\\&\quad{}
+\frac{i}{4\pi \lambda Z}
\log\frac{\big(\sqrt{A(Z-B)}-\sqrt{B(Z-A)}\big)^2}
{Z^2\left(\sqrt{Z-A}-\sqrt{Z-B}\right)^2}\,.
\eal
This expression still depends on $A$ and $B$, but there is no  easy
way to solve \eqn{omega0} and \eqn{omega1} to simplify this further, 
and need to keep those two constraints in mind below.

\subsection{The eigenvalue density}
The  exponentiated eigenvalues $Z$ 
are supported on the interval $[A,B]$, and their density is proportional to the 
discontinuity in $\omega$. This comes from $\sqrt{(A-Z)(B-Z)}$, leading to
\bal
\label{rho(Z)}
\rho(Z)&=
-\frac{i }{2\pi^2\lambda Z} \arctan\frac{(\sqrt{B}-\sqrt{A})\sqrt{(Z-A)(B-Z)}}
{\sqrt{A} (B-Z)+\sqrt{B} (Z-A)}
\\&\quad{}
+\frac{\chi}{\pi^2\mu Z} 
\left[
\arctan\sqrt{\frac{(B-Z)(A+M_K)}{(Z-A)(B+M_K)}}
-\arctan\sqrt{\frac{(B-Z)(A+M_1)}{(Z-A)(B+M_1)}}
\right].
\eal

We can easily go 
back to the original variables $z$ by the relation $\rho(z)dz=\rho(Z)dZ=2\pi Z\rho(Z)dz$, 
and equation
\eqref{eq:exponentiated_variables}.
We furthermore return to the original mass variables $m_i$ by the replacement 
$\mu=m_K-m_1$. This all gives us the expression for $\rho(z)$ as
\bal
\rho(z)&=
-\frac{i}{\pi\lambda}
\arctan\frac{\sqrt{\sh(z-a)\sh(b-z)}}{\ch\left(\frac{a+b-2z}{2}\right)}
\\&\quad{}
+\frac{2\chi}{\pi(m_K-m_1)}
\left[
\arctan\sqrt{\frac{\sh(b-z)}{\sh(z-a)}\frac{\ch(a-m_K)}{\ch(b-m_K)}}
-\arctan\sqrt{\frac{\sh(b-z)}{\sh(z-a)}\frac{\ch(a-m_1)}{\ch(b-m_1)}}
\right],
\eal
where the interval endpoints $a,b$ may be fixed by the normalisation conditions, or, 
equivalently, from the condition on the asymptotic behaviour of the resolvent 
(equations \eqref{omega0} and \eqref{omega1}).

\subsection{Wilson loops}

The vacuum expectation value of Wilson loops \cite{Gaiotto:2007qi} 
around a big circle of the $\bS^3$ may be computed as \cite{Kapustin2010}
\beq
\label{eq:Wilson}
W=\bigg\langle \frac{1}{N} \sum_{i} e^{2 \pi z_i}\bigg\rangle
\quad\underset{N\rightarrow \infty}{\longrightarrow}\quad
\int dz \rho(z)e^{2\pi z}
\,.\eeq
It turns out that this is easier to evaluate in the exponentiated variable $Z$, 
and it then takes the form
\beq
W
=\frac{1}{C}\oint dZ Z \omega(Z)
=\frac{1}{C}\int_A^B dZ Z \rho(Z)\,.
\eeq

Carrying out this integration, we find
\beq
W=
-\frac{i\big(\sqrt{A}-\sqrt{B}\big)^2}{8\pi C\lambda}
-\frac{\chi}{2\pi C\mu}\left(M_K-M_1-\sqrt{(A+M_K)(B+M_K)}+\sqrt{(A+M_1)(B+M_1)}\right)
\eeq
or, expressed in the original variables
\bal
\label{W2}
W&=
-\frac{i}{8 \pi\lambda}e^{\pi(a+b)}\sh^2\left(\frac{b-a}{2}\right)
-\frac{\chi}{2\pi(m_K-m_1)}
\bigg[e^{\pi(m_K+m_1)}\sh(m_K-m_1)
\\&\quad{}
-e^{\frac{\pi(a+b)}{2}}\left(e^{\pi m_K}\sqrt{\ch(a-m_K)\ch(b-m_K)}
-e^{\pi m_1}\sqrt{\ch(a-m_1)\ch(b-m_1)}\right) \bigg].
\eal

\subsection{Symmetric distributions}
\label{sec:sym}

In the special case of a symmetric mass distribution and vanishing FI-parameter, the entire 
problem has reflection symmetry around the origin. This means 
that $a=-b$, and so, the equations simplify
\beq
\label{eq:dens_org}
\rho(z)
=-\frac{i}{\pi\lambda}\arctan \frac{\sqrt{\ch^2(b)-\ch^2(z)}}{\ch(z)}
+\frac{\chi}{\pi m} \arctan\frac{\sh(m)\sqrt{\ch^2(b)-\ch^2(z)}}
{\ch(z)\sqrt{\ch^2(b)+\sh^2(m)}}\,.
\eeq
The condition \eqn{omega0} now takes the form
\beq
\label{omega0sym}
\chi-1=\frac{i}{\pi\lambda}\log\frac{\ch(b)}{2}
+\frac{\chi}{\pi m}\log\frac{\sqrt{\ch^2(b)+\sh^2(m)}+\sh(m)}{\ch(b)}\,,
\eeq
and we can use it to fix $b(m,\lambda,\chi)$. \eqn{omega1} is then 
automatically satisfied.

The Wilson loop \eqn{W2} is
\beq
W=
-\frac{i}{8 \pi \lambda }\sh^2(b)
-\frac{\chi}{4\pi m}\sh(m)
\left(\ch(m)-\sqrt{\ch^2(b)+\sh^2(m)}\right).
\eeq

\if0
In the case of vanishing $\chi$, so no fundamental fields, we find
\beq
\pi i\lambda=\log\cosh(\pi b)\,,
\eeq
while for vanishing CS level, $\lambda\to\infty$ and we find
\beq
\frac{\sqrt{\ch^2(b)+\sh^2(m)}+\sh(m)}{\ch(b)}
=e^{\pi m(1-1/\chi)}\,.
\eeq
\fi

\section{Form of the solution}
\label{sec:graphs}

In this section we study the different forms the eigenvalue density can take. We plot the 
analytic expressions for varying values of the parameters and compare it to a numerical 
analysis of the solution to the saddle point equations with $N=100$. The analytic expressions 
provide a good fit to the numerical results for the entire range of the parameters we tested 
(with negative imaginary 't~Hooft coupling).

\psfrag{rho}[][][1]{$\rho(z)$}
\psfrag{z}[][][1]{$z$}
\psfrag{-10}[][][.85]{$-10$}
\psfrag{-5}[][][.85]{$-5$}
\psfrag{-2.0}[][][.85]{$-2.0$}
\psfrag{-1.5}[][][.85]{$-1.5$}
\psfrag{-1.0}[][][.85]{$-1.0$}
\psfrag{-0.5}[][][.85]{$-0.5$}
\psfrag{-0.4}[][][.85]{$-0.4$}
\psfrag{-0.2}[][][.85]{$-0.2$}
\psfrag{0.5}[][][.85]{$0.5$}
\psfrag{1.0}[][][.85]{$1.0$}
\psfrag{1.5}[][][.85]{$1.5$}
\psfrag{2.0}[][][.85]{$2.0$}
\psfrag{2.5}[][][.85]{$2.5$}
\psfrag{3.0}[][][.85]{$3.0$}
\psfrag{5}[][][.85]{$5$}
\psfrag{10}[][][.85]{$10$}
\psfrag{0.01}[][][.85]{$0.01$}
\psfrag{0.02}[][][.85]{$0.02$}
\psfrag{0.03}[][][.85]{$0.03$}
\psfrag{0.04}[][][.85]{$0.04$}
\psfrag{0.05}[][][.85]{$0.05$}
\psfrag{0.1}[][][.85]{$0.1$}
\psfrag{0.2}[][][.85]{$0.2$}
\psfrag{0.3}[][][.85]{$0.3$}
\psfrag{0.4}[][][.85]{$0.4$}
\psfrag{0.6}[][][.85]{$0.6$}
\psfrag{0.7}[][][.85]{$0.7$}
\psfrag{0.8}[][][.85]{$0.8$}
\psfrag{chi=1}[][][1]{$\chi=1$}
\psfrag{chi=2}[][][1]{$\chi=2$}
\psfrag{k=0}[][][1]{$\kappa=0$}
\psfrag{mu=20}[][][1]{$\mu=20$}
\psfrag{mu=2}[][][1]{$\mu=2$}
\psfrag{mu=0.2}[][][1]{$\mu=0.2$}
\psfrag{l=-10i}[][][1]{$\lambda=-10i$}
\psfrag{l=-i}[][][1]{$\lambda=-i$}
\psfrag{chi}[][][1]{$\chi$}
\psfrag{lambda}[][][1]{$\tilde\lambda$}
\psfrag{m}[][][1]{$m$}
\psfrag{0}[][][.85]{$0$}
\psfrag{1}[][][.85]{$1$}
\psfrag{2}[][][.85]{$2$}
\psfrag{3}[][][.85]{$3$}
\psfrag{4}[][][.85]{$4$}
\psfrag{I}[][][.85]{$I$}
\psfrag{II}[][][.85]{$II$}
\psfrag{III}[][][.85]{$III$}
\psfrag{IV}[][][.85]{$IV$}
\begin{figure}[!ht]
\centering
\includegraphics[width=.45\textwidth]{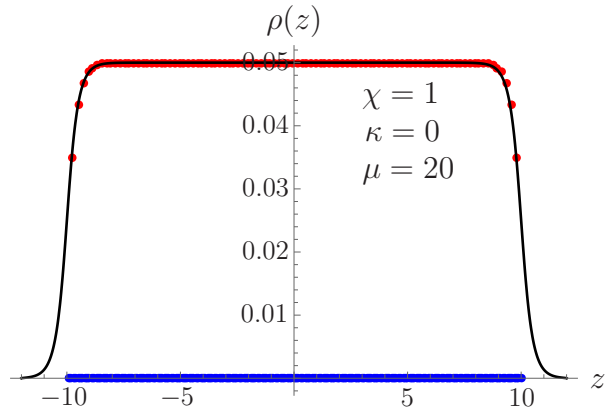}
\caption{The eigenvalue density for vanishing CS-level, $K=2N$ ($\chi=1$) and 
a relatively large $\mu=20$. 
The black line is the analytic density, and the red dots are numerical results for 
$N=100$. The blue dots are the distribution of masses, here condensed to a line.}
\label{fig:1}
\end{figure}

Figure~\ref{fig:1} shows the regime of large $\mu$, where the eigenvalue density 
approaches a constant. In this case, $m_K-m_1=20$ is not very large, so we see 
a few eigenvalues along the tails extending beyond the range of the masses 
$[-10,10]$. The limit of large $\mu$ is studied in detail in the next section.

\begin{figure}[!ht]
\centering
\includegraphics[width=.45\textwidth]{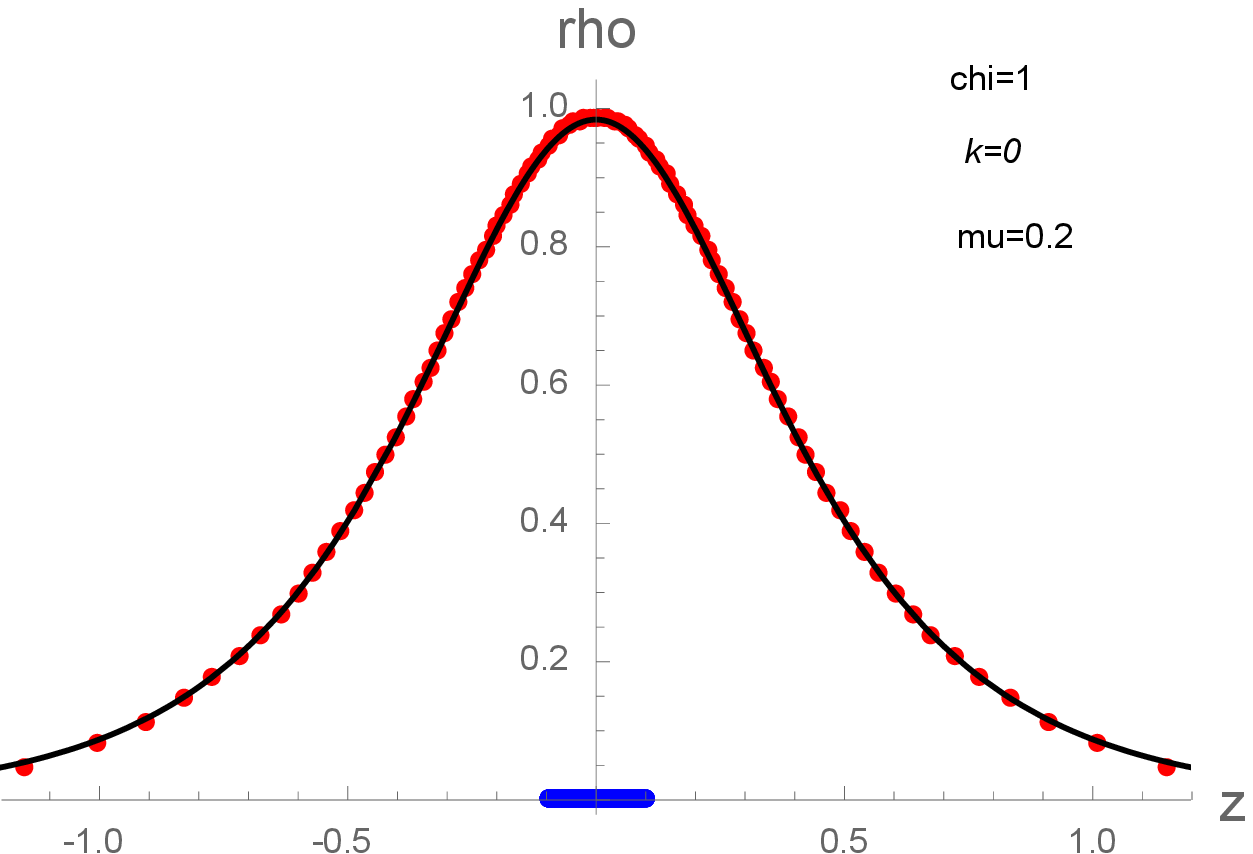}
\quad
\raisebox{.6mm}{\includegraphics[width=.45\textwidth]{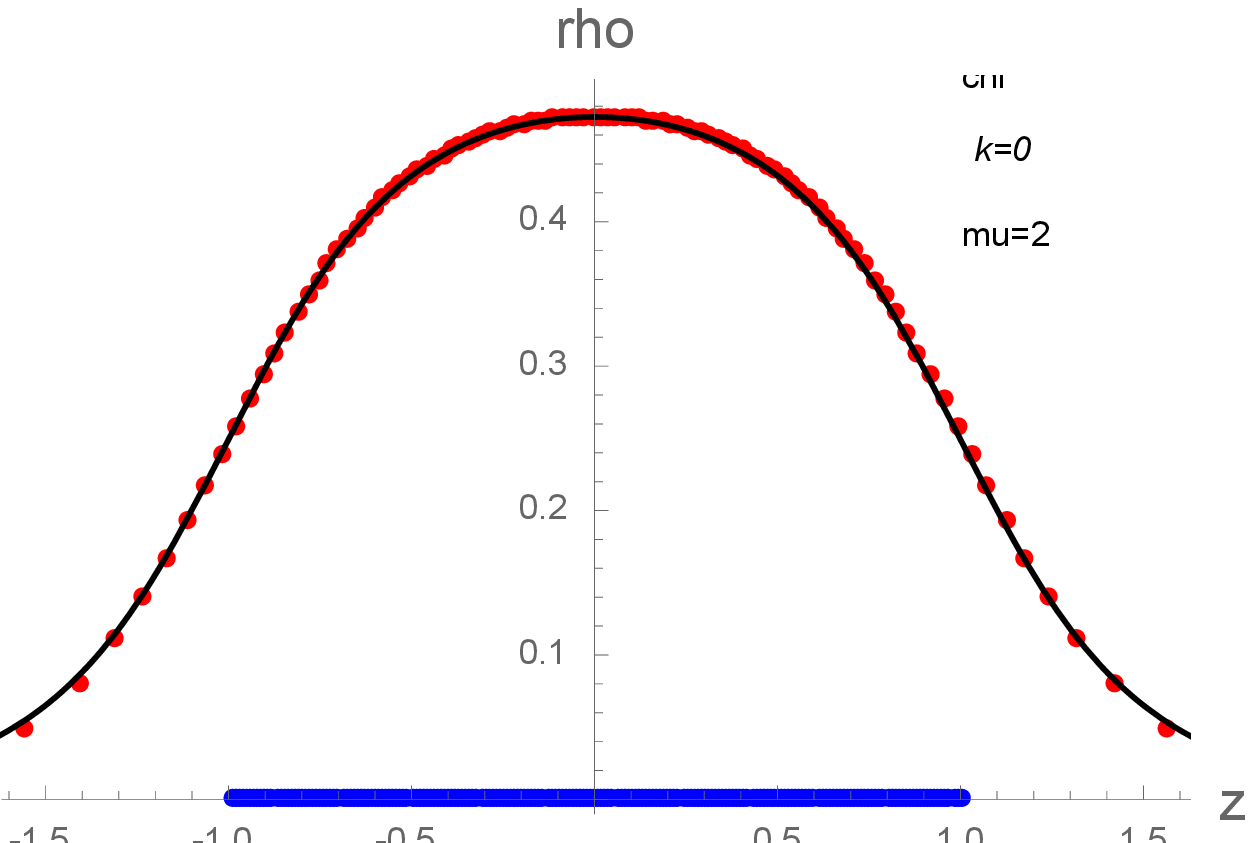}}
\caption{The eigenvalue density for vanishing CS-level, $K=2N$ ($\chi=1$) and 
varying $\mu$ smaller than in Figure~\ref{fig:1}.}
\label{fig:mu}
\end{figure}

When $\mu$ is small the eigenvalue distribution forms a bell shape, as can be seen in the left 
graph in Figure~\ref{fig:mu}. As we increase $\mu$, as in the right graph, the bell widens 
until it becomes flat, like in Figure~\ref{fig:1}.

\begin{figure}[!ht]
\centering
\includegraphics[width=.45\textwidth]{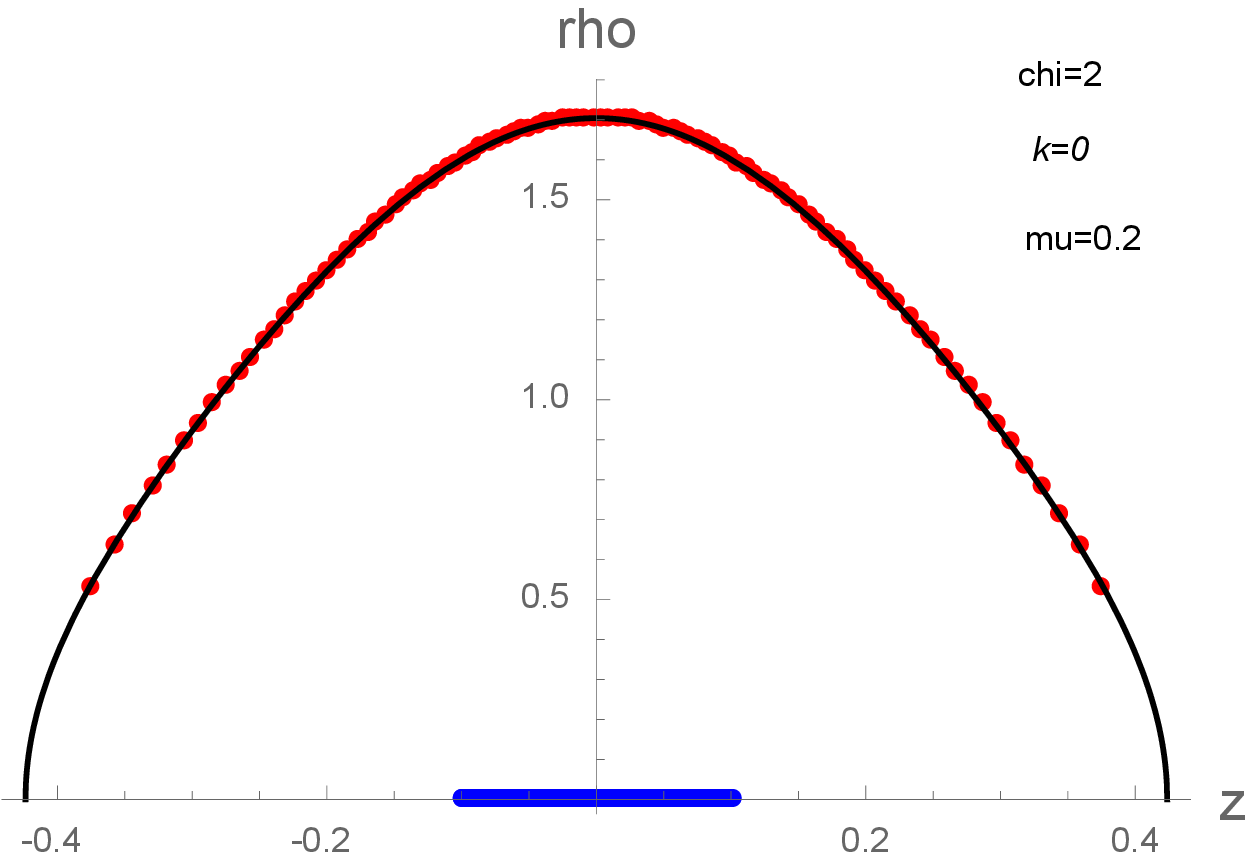}
\quad
\raisebox{.6mm}{\includegraphics[width=.45\textwidth]{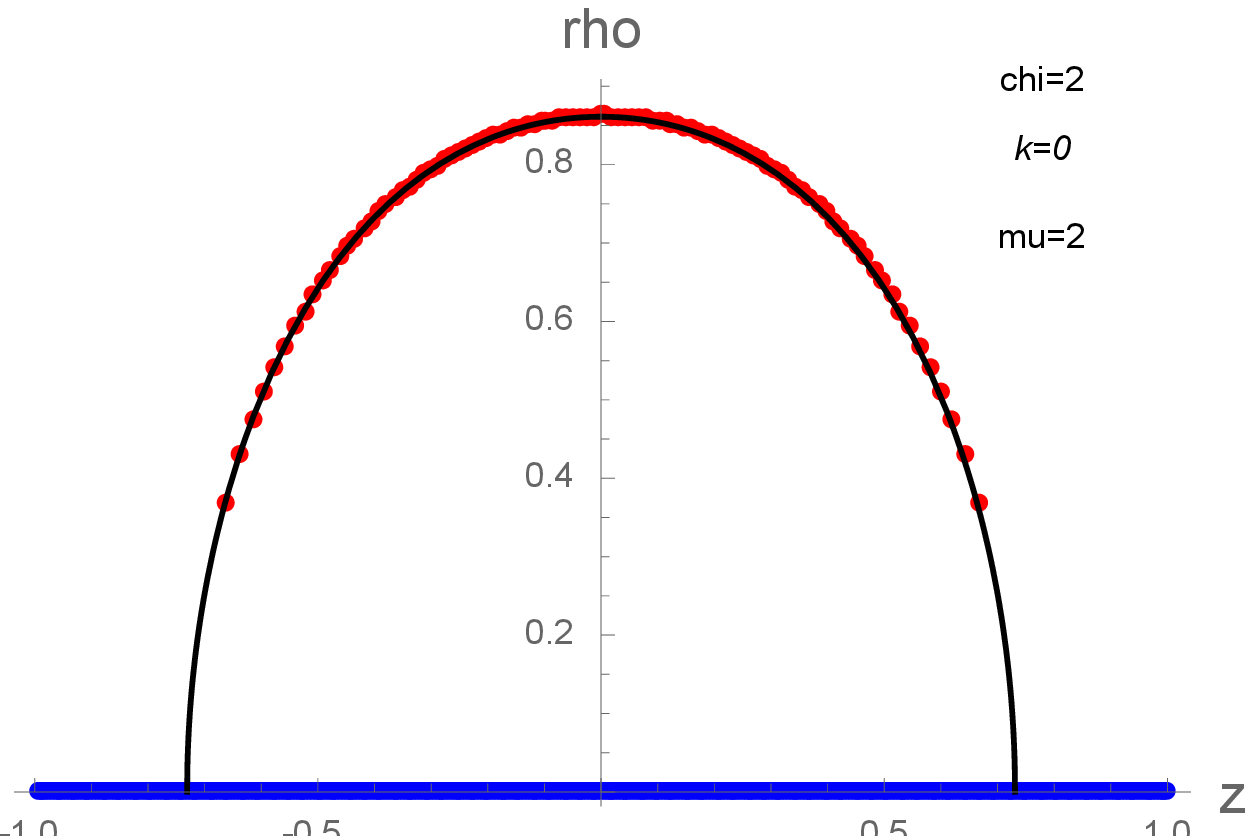}}
\caption{The eigenvalue density for vanishing CS-level, $K=4N$ ($\chi=2$) and 
the same two values of $\mu$ as in Figure~\ref{fig:mu}.}
\label{fig:chi}
\end{figure}

The eignevalues get squeezed further towards the origin if we increase $\chi$ (or $K$). 
In the absence of a CS term, the matrix model does not converge for $\chi<1$, but it can 
increase arbitrarily. In Figure~\ref{fig:chi} we illustrate the case of $\chi=2$ with the same 
values of $\mu$ as in Figure~\ref{fig:mu}. We see that with large enough $\mu$ and/or $\chi$, 
the eigenvalues no longer extend beyond the range of the mass distribution.

\begin{figure}[!ht]
\centering
\includegraphics[width=.45\textwidth]{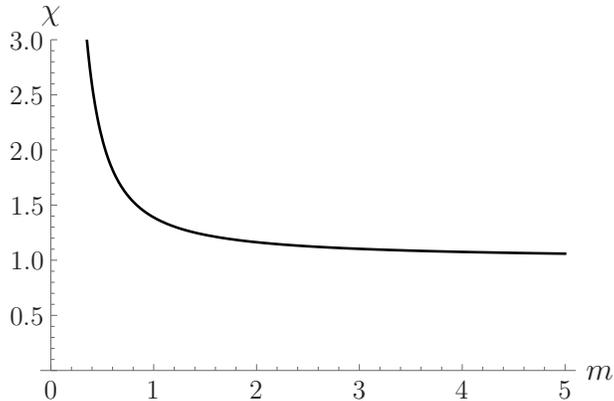}
\caption{The values of $\chi$ for which $b=m_K$ for symmetric mass distributions and 
$\kappa=0$.}
\label{fig:coincident}
\end{figure}

To check when this happens, we can solve for $b=m_K$, which also fixes $a=m_1$, 
as we are in the symmetric case, discussed in Section~\ref{sec:sym}. Equation 
\eqn{omega0sym} then gives
\beq
\label{b=m}
\chi-1=\frac{i}{\pi\lambda}\log\frac{\ch(m)}{2}
+\frac{\chi}{\pi m}\log\frac{\sqrt{2\ch(2m)}+\sh(m)}{\ch(m)}\,,
\eeq
In particular, if we focus on the case of $\kappa=0$, as in the above examples, we find 
a simple curve of $\chi(m)$, shown in Figure~\ref{fig:coincident}. It is easy to see that 
there are no discontinuities when crossing this line, so it does not lead to a phase transition. 
Those 
arise in the next section, when we consider the decompactification limit.

\begin{figure}[!ht]
\centering
\raisebox{.1mm}{\includegraphics[width=.45\textwidth]{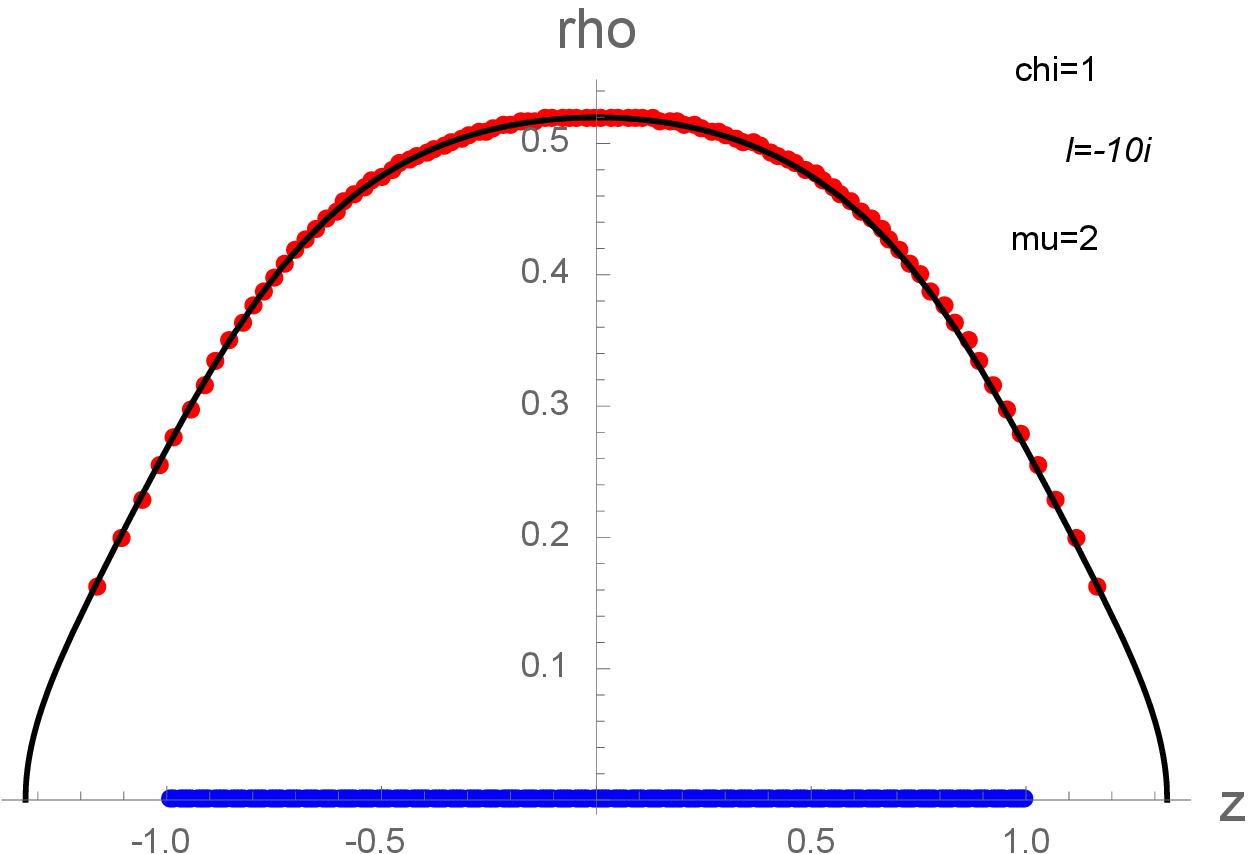}}
\quad
\includegraphics[width=.45\textwidth]{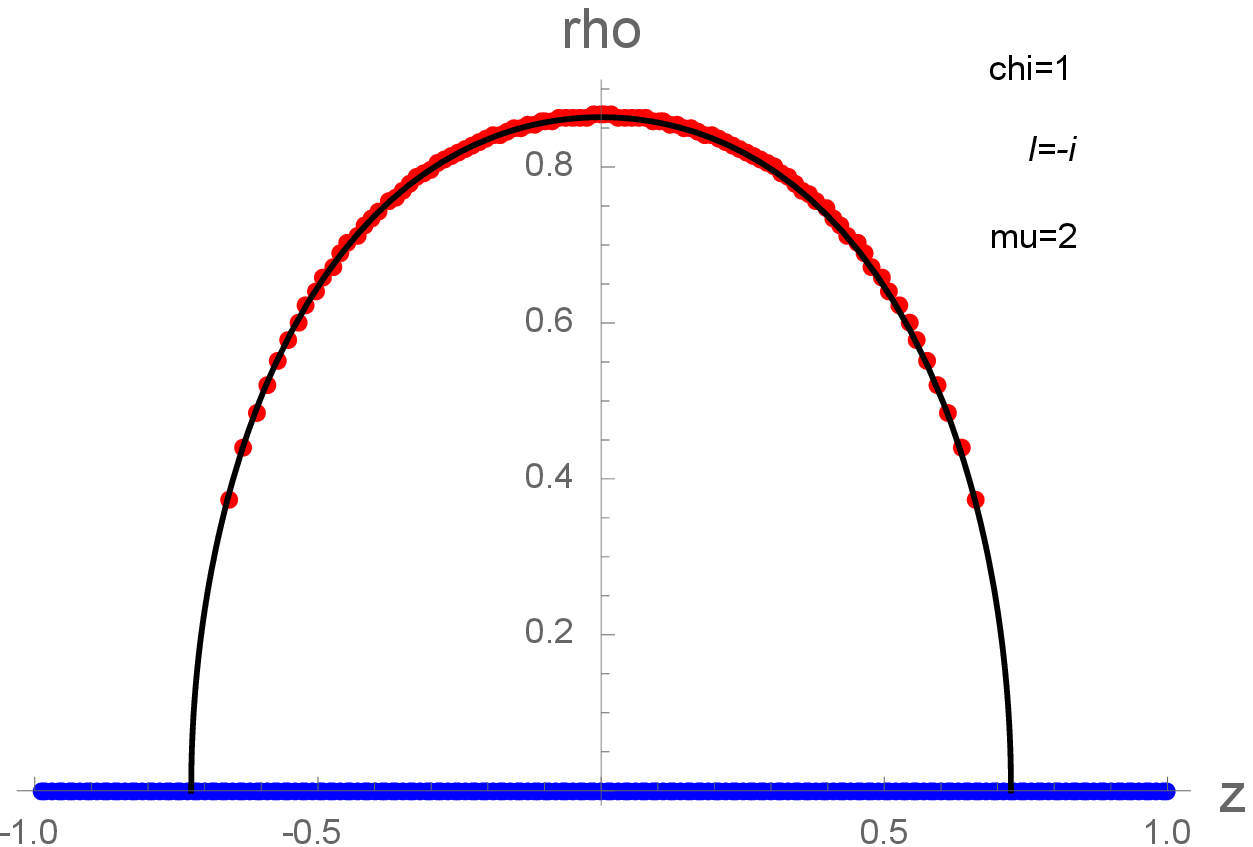}
\caption{Eigenvalue densities with varying CS-level.}
\label{fig:CS}
\end{figure}

The Chern-Simons coupling 
adds an extra potential term pushing the eigenvalues towards 
the origin, and in fact with a nonzero CS parameter 
we can take $\chi\to0$. Some plots with 
imaginary CS terms are shown in Figure~\ref{fig:CS}. 

The condition that the endpoints of the eigenvalue distribution coincide with the endpoints of 
the masses  \eqn{b=m}
now leads to a hypersurface in 3d.

\begin{figure}[!ht]
\centering
\includegraphics[width=.45\textwidth]{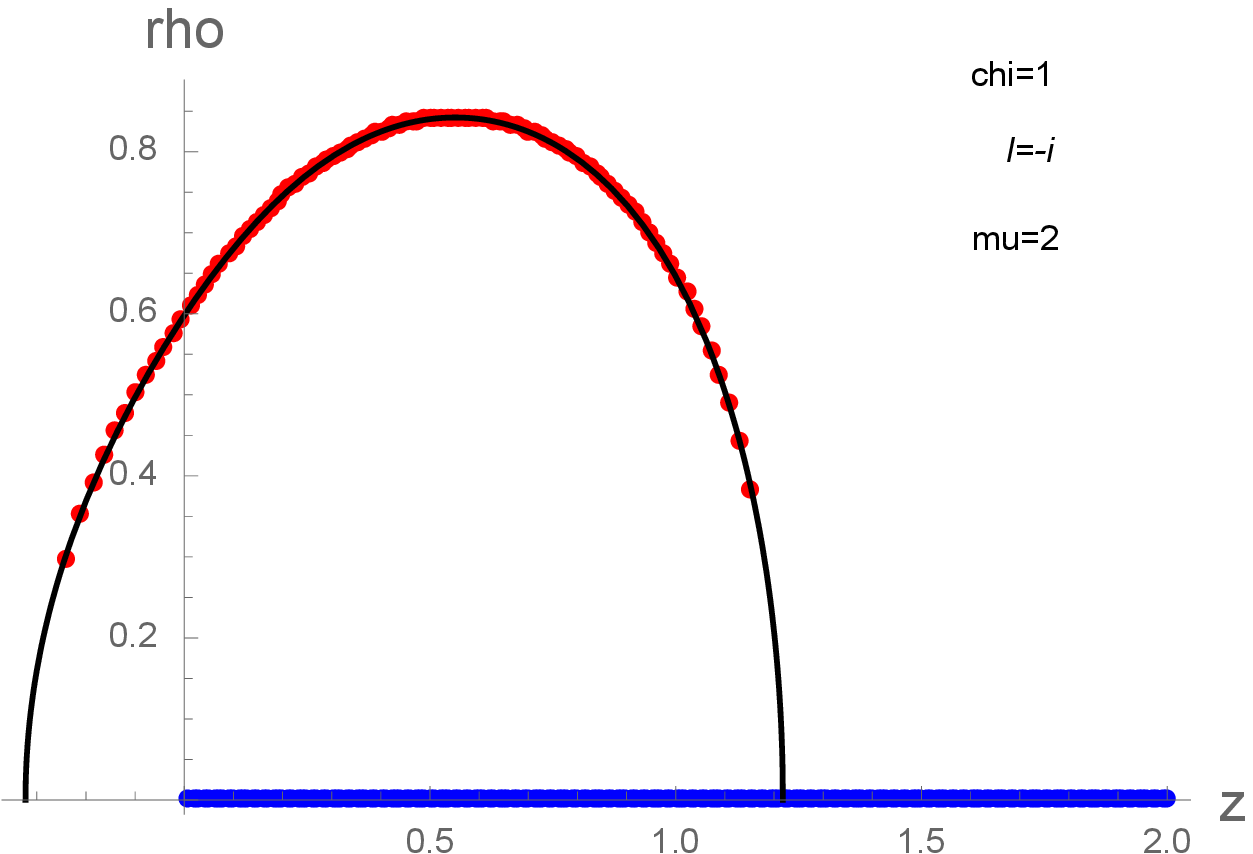}
\quad
\includegraphics[width=.45\textwidth]{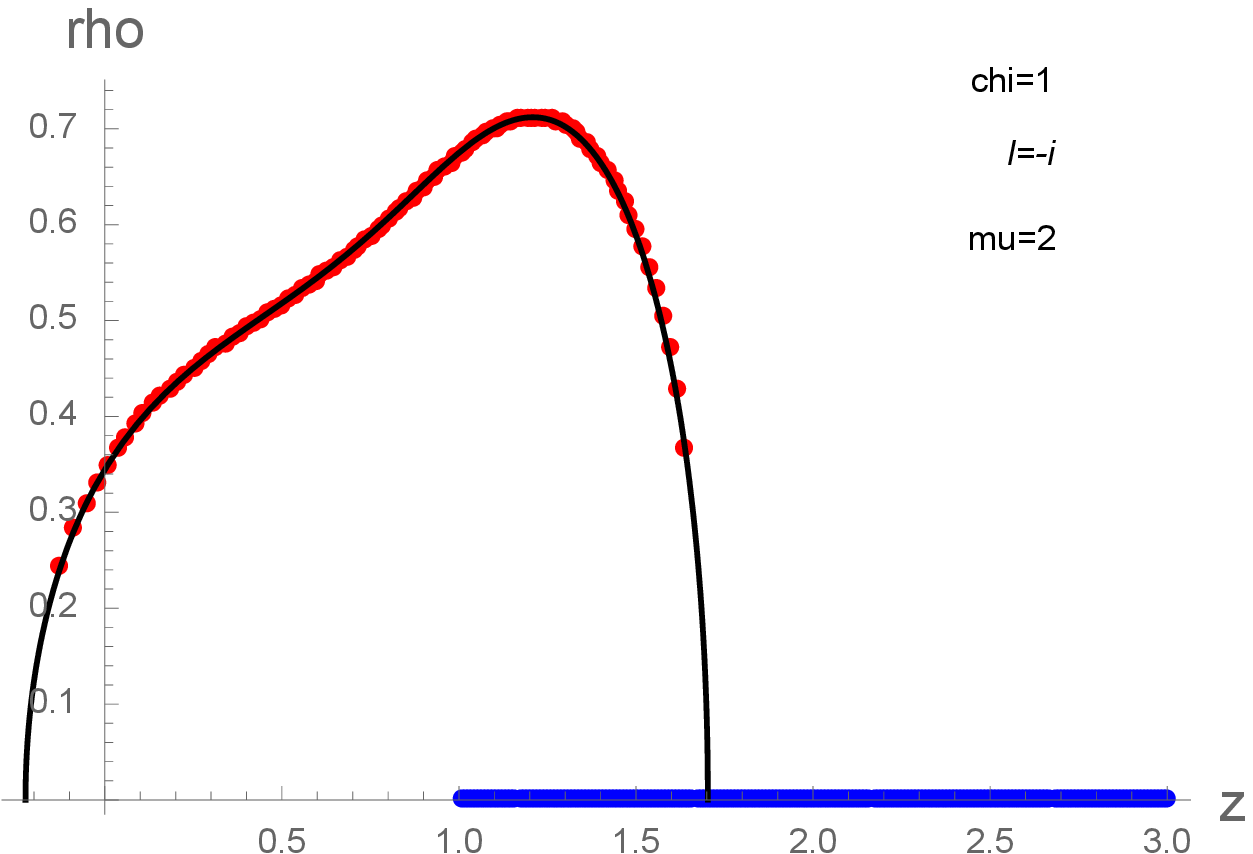}
\caption{Asymmetrical eigenvalue densities arising in the presence of 
nonzero CS-level and a mass distribution shifted 
relative to the right graph in Figure~\ref{fig:CS} by $1$ (left) and $2$ (right).}
\label{fig:asym}
\end{figure}

The situation gets more complicated when the mass distribution is not centered around the 
origin. In the absence of a CS term, the eigenvalue distribution translates with the masses, 
but the CS term breaks translation symmetry, and tends to keep the eigenvalues centred 
around the origin. Two graphs illustrating this are shown in Figure~\ref{fig:asym}. One can see 
that it may now be possible for one endpoint of the eigenvalue distributions to be outside of the 
domain of the masses, while the other is still inside. Depending on the range of the parameters, 
any order among $a$, $b$, $m_1$ and $m_K$ consistent with $a<b$ and $m_1<m_K$ 
is possible. Indeed, increasing the value of $m_1$
a bit beyond that in the right graph in 
Figure~\ref{fig:asym} leads to the eigenvalues completely disjoint from the mass 
distribution: $a<b<m_1<m_K$.

\section{The decompactification limit and phase transitions}
\label{sec:decomp}

Non-trivial phase structures has been observed previously for 
mass-deformed theories in 3, 4 and 5 dimensions 
\cite{Barranco:2014tla, Russo:2014bda, Anderson:2014hxa, 
Marmiroli:2014ssa, Zarembo:2014ooa, Russo:2014nka, 
Chen-Lin:2015dfa, Anderson:2015ioa, Nedelin:2015mta, Russo:2016ueu}.
It stands to reason that 
something similar might occur here. Therefore, we would like to examine 
the behaviour of the free energy and Wilson loops in the so-called 
decompactification limit where we take the radius $R$ of the $\bS^3$ to 
infinity while keeping $\kappa$ and $\lambda/R$ fixed. To that end we  
reintroduce the explicit radius dependance in the original matrix model 
\eqn{original-MM} (in which the radius of the sphere was taken to be unity) 
via the substitution
\beq
\label{eq:rescale}
m_j \rightarrow Rm_j\,,
\qquad
z_j \rightarrow Rz_j\,.
\eeq

The saddle point equation for the eigenvalue $z_i$ (with $\zeta=0$) now takes 
the form
\beq
\frac{1}{N}\sum_{ j \neq i}\coth(\pi R(z_i-z_j))
= \frac{\chi}{K}\sum_{k=1}^K \tanh(\pi R(z_i -m_k))- \frac{i}{\lambda}Rz_i\,.
\eeq
In the limit $R\rightarrow\infty$, the hyperbolic functions 
approach 
sign-functions of the real part of the argument. Let us introduce 
$\tilde{\lambda}=i\frac{\lambda}{R}$ which is held 
fixed (and positive) as $R\rightarrow \infty$. Going to the continuum limit, 
where the masses are distributed according to $\rho_m(m)$, we then find
\beq
\label{eq:saddle_largeR1}
\int_{a}^{b}dz'\,\rho(z')\sign(z-z')
=
\chi\int_{m_1}^{m_K}dm\,\rho_m(m)\sign(z -m)
+\frac{1}{\tilde{\lambda}}z\,.
\eeq
Differentiating both sides with respect to $z$  gives us the relation
\beq
\label{eq:saddle_largeR2}
\rho(z)=\chi \rho_m(z)+\frac{1}{2\tilde{\lambda}}\,.
\eeq
This is a remarkably simple expression for the eigenvalue density, 
which is valid for arbitrary mass distributions $\rho_m$, and not only 
the constant one which we analyzed previously for finite $R$. This equation 
is not enough, though, to determine the endpoints $a$ and $b$.

To that end, consider \eqn{eq:saddle_largeR1} for $z=a$ and $z=b$
\beq
\label{eq:saddle-end}
-1=2\chi\int_{m_1}^{a}dm\,\rho_m(m)-\chi+\frac{1}{\tilde{\lambda}}a\,,
\qquad
1=2\chi\int_{m_1}^{b}dm\,\rho_m(m)-\chi+\frac{1}{\tilde{\lambda}}b\,.
\eeq
We are interested in phase transitions as we change the parameter 
$\chi$ and coupling $\lambda$. These arise when $\rho(z)$ changes 
discontinuously and it is easy to see that for continuous $\rho_m$, the solution to 
\eqn{eq:saddle_largeR2} subject to the constraints in 
\eqn{eq:saddle-end} will depend continuously on $\chi$ and $\lambda$.

Phase transitions arise then from discontinuities in $\rho_m$, and focusing 
again on the case of constant mass-density, the discontinuities are 
at the endpoints $m_1$ and $m_K$. Given that $m_1<m_K$ and 
$a<b$, there are six possible arrangements of these variables.

\subsection{Symmetric mass distributions}
Before analyzing the general case, let us focus on 
symmetric mass distributions $m_K=-m_1=m$, which imply also 
$a=-b$. In this case there are two possible phases with 
$b<m$ and $b>m$. It is clear from \eqn{eq:saddle-end}, that the 
interface is at
\beq
1=\chi+\frac{m}{\tilde\lambda}\,.
\eeq
This matches the large $R$ limit of \eqn{b=m}, where the last term drops out. 
Note that since $\tilde\lambda>0$, this phase transition only occurs for $\chi<1$.

The case of $b<m$ leads to a constant eigenvalue density 
$\rho(z)=\frac{\chi}{2m}+\frac{1}{2\tilde\lambda}$ on the interval $[-b,b]$ 
and clearly $b=\frac{1}{2\rho(0)}$. This happens in the 
domain where $\tilde\lambda<\frac{m}{1-\chi}$.

For $b>m$, or $\tilde\lambda>\frac{m}{1-\chi}$ the eigenvalue 
density is
\beq
\rho(z)=\begin{cases}
\frac{1}{2\tilde\lambda}+\frac{\chi}{2m}\,,\qquad\quad&
-m<z<m\,,
\\
\frac{1}{2\tilde\lambda}\,,&
m<|z|<b\,,
\\
0\,,&|z|>b\,,
\end{cases}
\eeq
and $b=\tilde\lambda(1-\chi)$.

We can evaluate the free energy on both sides of the phase transition. For a given 
$\rho_m(m)$ and $\rho(z)$ it is given in this limit by
\bal
F
=\log N!-\pi N^2R\left[\int dzdz'\, \rho(z)\rho(z')|z-z'|
-
2\chi\int dzdm\, \rho(z)\rho_m(m)|z-m|
-\frac{1}{\tilde\lambda}\int dz\,\rho(z)z^2\right].
\eal
We thus find
\beq
F=\log N!
+\begin{cases}
\pi N^2R\left(m\chi-\frac{m\lambda}{3(m+\lambda\chi)}\right)
,
\qquad&\tilde\lambda<\frac{m}{1-\chi}\,,
\\
\pi N^2R\left(m\chi-\frac{m\lambda}{3(m+\lambda\chi)}
-\frac{\chi}{3\lambda(\lambda\chi+m)}(\lambda(1-\chi)-m)^3\right)
\,,
\qquad&\tilde\lambda>\frac{m}{1-\chi}\,.
\end{cases}
\eeq
Clearly this is continuous at $b=m$, and so are the first and second 
derivatives with respect to $\tilde\lambda$ or $\chi$ or $m$, so this is a third order 
phase transition.

Similarly, the Wilson loop may be computed by \eqref{eq:Wilson}, giving
\beq
W=
\begin{cases}
\frac{(\tilde{\lambda}\chi+m)}{4\pi\tilde{\lambda} mR}
\sh\left(\frac{2\tilde{\lambda}mR}{\tilde{\lambda}\chi+m}\right)
\,,
\qquad& \tilde\lambda<\frac{m}{1-\chi}\,,
\\
\frac{\chi}{4\pi mR}\sh(2mR)
+\frac{1}{4\pi\tilde{\lambda}R}\sh\left(2\tilde{\lambda}(1-\chi)R\right)
\,,
\qquad& \tilde\lambda>\frac{m}{1-\chi}\,.
\end{cases}
\eeq
Unlike the free energy, this has a discontinuity already in the second derivative. 
Since we are in the large $R$ limit we can replace all $\sh$ functions with 
$\exp$ and away from the phase boundary can also ignore the subleading exponent 
in the second phase, so

\beq
W\simeq
\begin{cases}
\frac{(\tilde{\lambda}\chi+m)}{4\pi\tilde{\lambda}mR}
e^{2\pi\tilde\lambda mR/(\tilde\lambda\chi+m)}
\,,
\qquad& \tilde\lambda<\frac{m}{1-\chi}\,,
\\
\frac{ \chi}{4\pi mR}e^{2\pi mR}
\,,
\qquad& \tilde\lambda>\frac{m}{1-\chi}\,,
\end{cases}
\eeq
which, just as the free energy, is continuous at $b=m$, but the second derivative with 
respect to $\tilde\lambda$, $\chi$ or $m$ is not.

\subsection{Asymmetric mass distribution}

In the generic case there are six phases which are most easily classified by 
the arrangement of $a$, $b$, $m_1$ and $m_K$, as listed in Table~\ref{tab1}. 
Of course $a$ and $b$ depend on the values of the parameters $\chi$, 
$\tilde\lambda$ and the masses, which is also illustrated in Table~\ref{tab1}.

\begin{table}[!ht]
\begin{center}
\begin{tabular}{|>{$}c<{$}|>{$}c<{$}|>{$}c<{$}|>{$}c<{$}|}
\hline
&&&\\[-4mm]
\text{Phase}&\text{Configuration} & a& b
\\[1mm]
\hline
&&&\\[-4.5mm]
\hline
&&&\\[-4mm]
I&
a<b< m_1< m_K
& (\chi-1)\tilde\lambda
& (\chi+1)\tilde\lambda
\\[1mm]
\hline
&&&\\[-4mm]
II&
a< m_1<b< m_K
&
(\chi-1)\tilde\lambda
& 
\frac{\tilde\lambda(m_1(\chi-1)+m_K(\chi+1))}{2\tilde\lambda\chi+m_K-m_1}
\\[1mm]
\hline
&&&\\[-4mm]
III&
a<m_1<m_K<b 
& (\chi-1)\tilde\lambda
&
(1-\chi)\tilde\lambda
\\[1mm]
\hline
&&&\\[-4.5mm]
IV&
m_1<a<b<m_K 
& 
\frac{\tilde\lambda(m_1(\chi+1)+m_K(\chi-1))}{2\tilde\lambda\chi+m_K-m_1}
& 
\frac{\tilde\lambda(m_1(\chi-1)+m_K(\chi+1))}{2\tilde\lambda\chi+m_K-m_1}
\\[1.5mm]
\hline
&&&\\[-4.5mm]
II'&
m_1<a<m_K<b
& 
\frac{\tilde\lambda(m_1(\chi+1)+m_K(\chi-1))}{2\tilde\lambda\chi+m_K-m_1}
& 
(1-\chi)\tilde\lambda
\\[1.5mm]
\hline
&&&\\[-4mm]
I'&
m_1<m_K<a<b 
&
-(\chi+1)\tilde\lambda
&
(1-\chi)\tilde\lambda
\\[1mm]
\hline
\end{tabular}
\end{center}
\vskip-4mm
\caption {Values of $a$ and $b$ in all phases. Notice that 
not all phases are possible for all values of the masses.}
\label{tab1}
\end{table}

\begin{figure}[!ht]
\centering
\includegraphics[width=.45\textwidth]{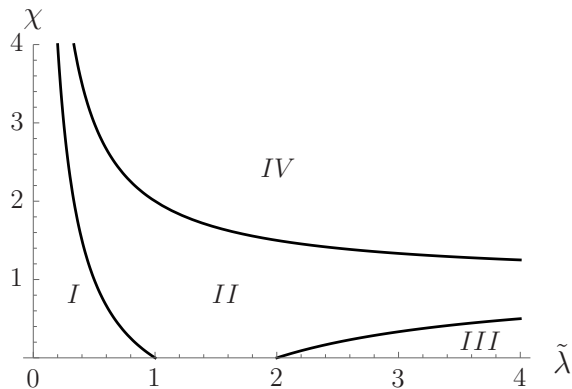}
\caption{A phase diagram for $m_1=1$ and $m_K=2$ where  phases 
$I$, $II$, $III$ and $IV$ are realized. 
The $I-II$ boundary curve is $1=(\chi+1)\tilde\lambda$, 
the $II-IV$ one is given by $1=(\chi-1)\tilde\lambda$ and the $II-III$ one by 
$2=(1-\chi)\tilde\lambda$.}
\label{fig:phases}
\end{figure}

If we fix $m_1$, $m_K$ then at most four phases can be seen by changing 
$\chi$ and $\lambda$. 
The phase boundaries are at
\beq
b=m_1\,,\quad
b=m_K\,,\quad
a=m_1\,,\quad
a=m_K\,.
\eeq
The expressions for the different phase boundaries are given in Table~\ref{tab2} and 
the phase diagram in the $(\tilde\lambda,\chi)$-plane for $m_1=1$ and $m_K=2$ 
is shown in Figure~\ref{fig:phases}.
\begin{table}[!ht]
\begin{center}
\begin{tabular}{|>{$}c<{$}|>{$}c<{$}|>{$}c<{$}|}
\hline
&&\\[-4.5mm]
\text{Phase-boundary}&\text{Configuration} & \text{condition} 
\\[1mm]
\hline
&&\\[-4.5mm]
\hline
&&\\[-4.5mm]
I-II&
a<b= m_1< m_K 
& m_1=(\chi+1)\tilde\lambda
\\[1mm]
\hline
&&\\[-4.5mm]
II-III&
a<m_1<b= m_K 
& m_K=(1-\chi)\tilde\lambda
\\[1mm]
\hline
&&\\[-4.5mm]
II-IV&
a=m_1<b< m_K 
& m_1=(\chi-1)\tilde\lambda
\\[1mm]
\hline
&&\\[-4.5mm]
III-II'&
a=m_1< m_K<b
& m_1=(\chi-1)\tilde\lambda
\\[1mm]
\hline
&&\\[-4.5mm]
IV-II'&
m_1<a<b= m_K 
& m_K=(1-\chi)\tilde\lambda
\\[1mm]
\hline
&&\\[-4.5mm]
II'-I'&
m_1< m_K=a<b
& m_K=-(\chi+1)\tilde\lambda
\\[1mm]
\hline
\end{tabular}
\end{center}
\vskip-4.5mm
\caption {The phase boundaries.}
\label{tab2}
\end{table}

It is easy to evaluate the free energy and the Wilson loop 
in each one of the phases, following the example in the symmetric case, 
and one finds once again all the phase transitions to be of third order.

\section{Discussion}
\label{sec:discuss}

The purpose of this note was to study further large $N$ limits of matrix models which 
arise from localization of 3d supersymmetric field theories and hence solve for their 
$\bS^3$ partition function. The models we 
studied has $K$ fundamental fields, whose number scales with $N$ in the limit and 
the novel feature that we addressed is to allow different masses for the fundamental 
fields.

For even $K$ we can view our model as an ungauged version of ABJ theory with 
gauge group $U(N)\times U(K/2)$. In the ABJ model one would have to integrate 
over the Coulomb branch parameters of the $U(K/2)$ vector multiplets, but when 
ungauged, those get frozen into the mass distribution. An additional Vandermonde-like 
factor, arising from the one-loop determinant of the vectors (and the Haar measure), 
is also absent, but could easily be incorporated as a simple determinant of the 
mass parameters.

We chose the simplest nontrivial mass distribution, a constant distribution over an 
arbitrary domain $[m_1,m_K]$, and were able to solve the matrix model in the 
large $N$ limit. In fact, there is 
a limit of the ABJ model (and similar theories), where the eigenvalue distribution 
is approximately flat \cite{Herzog2011} and also in our case, we find such a solution, see 
Figure~\ref{fig:1}. Starting with the ABJ model in the regime where the eigenvalue 
distribution is flat and ungauging it, keeping one set of eigenvalues as dynamical 
and the other frozen, will lead to the same solution.

Therefore, we could start with the ABJ model in a different regime, where the eigenvalue 
distribution is not flat, ungauge one group and we are guaranteed that the solution for 
the other set of eigenvalues would not be modified. It would be interesting to explore 
others mass distributions arising in this or other ways.

We have studied the phase structure of the model and found that in the decompactification 
limit there are six different phases as one modifies the mass distribution, $K$ and the 
Chern-Simons coupling. For finite radius spheres we found no phase transitions, but 
we should emphasize that our solution assumed a single cut, and while this is consistent 
with the numerical tests, it cannot be seen as a conclusive statement. In particular 
we have worked exclusively with imaginary CS parameter, which helps convergence. Our 
solution can be analytically continued to real values of the CS 
parameter, but our tests of the phase structure may no longer be valid.

Another natural generalization of our model is the case of a longer linear quiver. Such 
models are again easy to solve at finite $N$ \cite{Benvenuti:2011ga}, and can also 
be written as ungauged circular quivers, which are also easy to solve at large $N$ 
\cite{Herzog2011,Marino2012}. We leave that to future work.

\section*{Acknowledgments}
We are grateful to Sara Pasquetti, Jorge Russo
and Kostya Zarembo for enlightening discussions. 
The work of L.A. is supported  by the EPSRC programme grant ``New Geometric
Structures from String Theory'', EP/K034456/1. 
The work of N.D. is supported by Science \& Technology Facilities Council 
via the consolidated grant number ST/J002798/1.

\bibliographystyle{utphys2}
\bibliography{references.bib}

\end{document}